\documentclass[aps,prx,twocolumn,groupedaddress]{revtex4-2}
\usepackage{bm, amssymb,amsmath}
\usepackage{graphicx}
\usepackage[utf8x]{inputenc}
\usepackage[dvipsnames]{xcolor}
\usepackage{hyperref}
\usepackage{svg}
\usepackage{float}
\hypersetup{colorlinks=true,linkcolor=WildStrawberry,filecolor=magenta,citecolor=ProcessBlue,urlcolor=ProcessBlue,pdftitle={Charge asymmetry in the Heisenberg model}}

\begin{document}


\title{}


\author{Rohit Hegde}
\email[rohegde@gmail.com]{}


\date{\today}

\begin{abstract}
Supplementing the Heisenberg model with a Hubbard-commuting kinetic of electrons adds to its spectrum without interference. One consequence is the precise incorporation of canonical linear spin wave theory within the time-dependent Hartree-Fock framework, as pure localization emerges from itinerant dynamics. This embedding method generalizes to all spin-½ models and is expected to extend to multi-orbital systems. Away from half-filling, differential tuning of doublon and holon motion imparts asymmetry to ordering and fluctuations. This suggests that, in effective electronic theories, kinetic interaction couplings are as significant as underlying band parameters when modeling asymmetric phenomena near the Mott insulator. 
\end{abstract}


\title{Charge asymmetry in the Heisenberg model}
\maketitle

Charge asymmetry typically discriminates between particles and holes, as seen in the Fermi liquid parameters effectively characterizing their movements and interactions. In the alternative Mott insulators and related quantum phases, doublons and holons ascend as fundamental charges either alongside or replacing particles and holes, imbuing a natural language for model dynamics. The layered cuprate materials, with their iconically asymmetric presentation of antiferromagnetism and superconductivity, and as candidate targets of strongly coupled Hubbard models, motivate a theory of electronic order that is effectively written in terms of doublon and holon dynamics.

This paper explicates such a theory with a unifying description of localized and itinerant phases in the electronic Hilbert space. It is shown that all localized spin models, exemplified by the Heisenberg Hamiltonian, can be purely embedded in the electronic Hilbert space alongside doublon and holon motion. In this framework, the kinetic energy incorporates two- and three-body interactions that extend beyond traditional one-body hopping. The purity of embedding reflects Hubbard-commutativity of kinetic energy and has significant implications for the Gaussian theory of order (HF) and fluctuations (TDHF). At the static level, a purely local density matrix emerges at half-band filling from dispersive quasiparticle bands.  In the time-dependent treatment, a spin-wave constitutes depending purely on the spin parameters $J_{ij}$ and not on the charged quantities $t_{ij}, u$, precisely subsuming canonical linear spin wave theory, and reconciling traditional methods fashioned for insulators with those developed for metals.  

\begin{figure}
    \centering
    \includegraphics[scale=0.4]{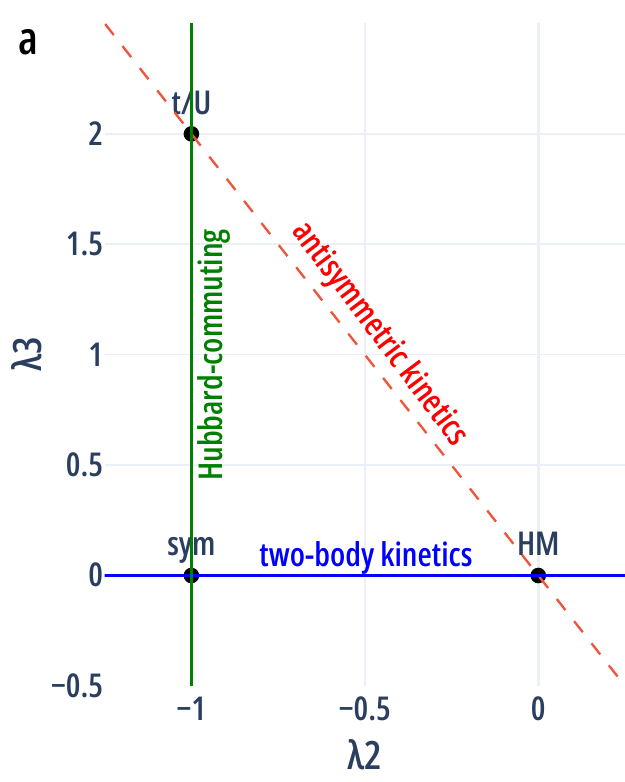}
    \includegraphics[scale=0.4]{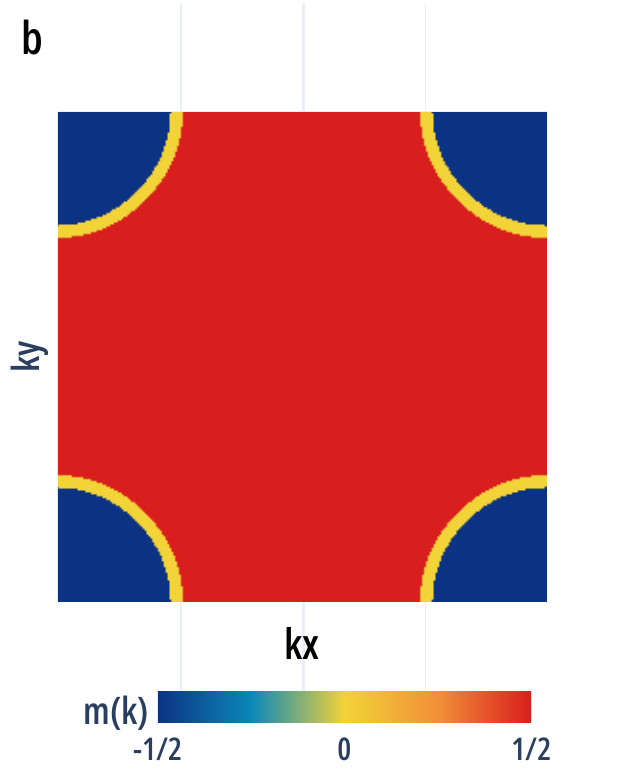}
    \caption{(\textbf{a})Promotion of band hopping to two-parameter$(\lambda_2,\lambda_3)$ interacting kinetic allows for the study of Hubbard model, t/U series, an $SU(2)$ charge symmetric model, and interpolating dynamics in terms of electronic fields. Canonical linear spin wave theory precisely emerges from time-dependent Hartree-Fock applied to the set of Hubbard-commuting dynamics, which have tunable charge asymmetry. (\textbf{b})The general model allows access to novel correlated states beyond the Mott insulator regime. A kinetic ferromagnet has $n_{\bm{k}}=1$ at half-filling, but is delocalized in charge, and primarily in spin, as the magnetization is a binary variable over $\bm{k}$. Doping the ferromagnet leaves a Fermi ring around the $M$ point.}
    \label{fig:1}
\end{figure}

The fundamental premise of this effective model is that the electronic interactions kinetically sustaining the Mott insulator—specifically, the two-body correlated hopping ($\lambda_2$) and three-body explicit doublon hopping ($\lambda_3$)—are ontologically real and capable of influencing phenomena, much like the lower-order Hubbard interaction and higher-order spin exchange. One test is in the model's expression of magnetism, in its dependency on the kinetic parameters ($\lambda_2,\lambda_3$). By studying dependence of the spiral pitch of the coplanar state on model parameters, it is found that $\lambda_2$ serves as a tightness parameter; essentially interpolating between the Hubbard and Heisenberg models.  Nearly Hubbard-commuting kinetics allow for the expression of large-U behavior at intermediate or even small U, by decreasing quasiparticle bandwidth, and sensitizing it to the form of spin order. Meanwhile $\lambda_3$ serves primarily as a conveyor and modulator of the underlying band asymmetry to the interacting dynamic. The sharpest distinction between interacting and noninteracting kinetic energies is seen in an exotic ferromagnet that is gently reminiscent of collinear states  intertwined with superconductors \cite{PhysRevLett.86.846}, or those emergent from strong spin-orbit coupling. The ferromagnet depicted in Fig \ref{fig:1} is representative of correlated but non-Mott insulator states that can only emerge from a dynamic able to foster Hubbard-commutativity, while suppressing the value of the Hubbard coupling, as a chemical potential for doublons and holons.

A practical advantage of the electronic construction of pure Mott localization is its applicability beyond half-filling without the unphysical degrees of freedom present in gauge mean-field theory.  Unlike the t-J model and projected dynamics, the present class of Hamiltonians can operate on states with varying numbers of doublons and holons, surpassing the constraints of the infinite U limit. Of potential use in comporting with phenomenology is the tunability of the embedding, allowing for a softer, and presumably more realistic Mott insulator than the ideal of pure symmetry. 

Three models serve as waypoints in the two-dimensional parametric plane under study.  Two familiar models, the Hubbard and the t/U Hamiltonians\cite{PhysRevB.37.9753}, are shown to be points on a line of charge antisymmetric kinetics, with a one-parameter unitary generating the line from the Hubbard model. A third, singularly charge symmetric dynamic has been previously identified for its $SU(2)$ charge symmetry and Hubbard-commutativity, and studied for the constraints placed by these symmetries on the spectrum\cite{PhysRevLett.74.789,PhysRevLett.68.2960,PhysRevLett.70.73}. As far as we know, neither the t/U, nor the symmetric model have been studied in a standard order parameter construction like the TDHF theory presented here.

Section \ref{sec2} introduces the model, focusing on the charge symmetry of the interacting kinetic and the relation to the Hubbard and t/U models. It is shown that the t/U dynamic is the sole Hubbard-commuting member of an antisymmetric family, and the one-parameter generalization of the unitary transformation is provided. The Hartree-Fock equation generated by the interacting model is solved in the context of Neel antiferromagnetism, demonstrating pure localization from itinerancy. Section \ref{sec3} makes clear the exact subsummation of canonical linear spin wave theory by time-dependent Hartree-Fock, built upon the dynamical emergence of extensively many local spin flip bosons, having zero energy, from the lower-order dynamic that lacks Heisenberg's term.  

\section{Localization from itinerancy}
The simplest way to effect localization in a noninteracting model is to cancel the kinetic energy and flatten the band. The remaining dynamical evolution keeps charge fully in place for all states acted upon. In a narrower, yet more nuanced occurrence, a particular local Slater determinant state is selectively annihilated by one or more finite kinetics.

Consider an arrangement of spins on a lattice, $\Phi_{\bm{\alpha}}=\prod_i c^\dagger_{i \bm{\alpha}_i}|\varnothing\rangle$, oriented in the directions $\{\bm{\alpha}_i\}$. This Slater determinant is annihilated by a pair of general hopping matrices $t^\pm_{ij}$ that describe the motion of $\pm\hat{z}$-components of spin in the comoving frame of $\Phi_{\bm{\alpha}}$, the frame in which the state is a uniform, $\hat{z}$-polarized ferromagnet. In that frame, localization is simply an effect of Pauli exclusion.  In the default frame, the one-body kinetic localizing $\Phi_{\bm{\alpha}}$ is spin-dependent,
\begin{align}
    &T^{(\alpha)}_{loc}= \sum_{ij}\sum_{s=\pm}t_{ij}^{s} c^\dagger_{i\, s \bm{\alpha}_{i}}c_{j \,s \bm{\alpha}_{j}}, \nonumber \\
    &T^{(\alpha)}_{loc} \Phi_{\bm{\alpha}}  =0, \;\;\;\;\;\;\forall t_{ij}^{\pm}.\label{Tlocal}
\end{align}
While \eqref{Tlocal} may appear finely tuned in its dependence on $\{\bm{\alpha}_i\}$, the localizing noninteracting kinetics are automatically generated by Wick contracting the set of interacting, Hubbard-commuting kinetics. The former are parameterized by two bands $t_{ij}^{\pm}$, the independence of which allows for particle-hole asymmetry, while the latter are characterized by a single band $t_{ij}$ and a coupling constant $\lambda_3$ that discriminates between doublon and holon hopping. In a mean-field theory, the values of $t_{ij}^{\pm}$ inherit from $t_{ij}$ and $\lambda_3$, and charge asymmetry passes from doublons and holons to quasiparticles and holes.

\section{Interacting model\label{sec2}}
Resolving noninteracting kinetic energy as a sum of interacting parts, $T=\sum_{ij\sigma}t_{ij}(n_{i\bar{\sigma}}+h_{i\bar{\sigma}})c^\dagger_{i\sigma}c_{j\sigma}(n_{j\bar{\sigma}}+h_{j\bar{\sigma}})$ shows the band hopping as the Hubbard interaction sees it,
\begin{align}
    &T_{\alpha,\beta}=T_h+\alpha T_d +\beta(T_+ + T_-) , \\
    &T=T_{1,1} \,, \nonumber
\end{align}
and provides a natural interacting deformation in terms of the couplings $\alpha,\beta$. Setting $\beta=0$ yields the set of Hubbard-commuting kinetics, as $T_{\pm}$ are the operations that annihilate doublon-holon pairs into singlons and vice versa.

Constitution of an electronic theory requires recasting in terms of normal-ordered operators, $T_{\alpha,\beta} \to T_{\lambda_2,\lambda_3}\,$,
\begin{align}
    &T_{\lambda_2,\lambda_3}=T+\lambda_2 T_2 + \lambda_3 T_3 , \\
    &T_2=\sum_{ij\sigma} t_{ij} c^\dagger_{i\sigma}c_{j\sigma}\, (n_{i\bar{\sigma}} +n_{j\bar{\sigma}}), \\
    &T_3=\sum_{ij\sigma} t_{ij} c^\dagger_{i\sigma}c_{j\sigma}\, n_{i\bar{\sigma}} \,n_{j\bar{\sigma}}   =T_d ,
\end{align}
so that $T=T_{0,0}\,$, and $T_{-1,\lambda_3}$ are the set of commuting kinetics.

By default, charge symmetry in this article refers to spin-commuting charge conjugation, $\mathcal{C}: c_\alpha \rightarrow \epsilon_{\alpha\beta} c_\beta^\dagger$, as opposed to the more nuanced and typical sense of combining conjugation with a sublattice-dependent sign change, $\mathcal{C}_G = \mathcal{C} (-1)^{x+y}$. Then, all band hoppings are antisymmetric, $\mathcal{C}^\dagger T \mathcal{C}=-T$, as is the encompassing, linear family of dynamics, $T_{-\eta,2\eta}$, which includes the kinetic of the t/U series given by $\eta=1$. The family can be generated from the Hubbard model by a one-parameter unitary transformation, as in $\mathcal{U}^\dagger_{\eta}(U+T)\mathcal{U}_{\eta}=U+T_{-\eta,2\eta}+h.o.t.$, with the higher order terms being non-uniquely determined except in the case of $\mathcal{U}_1$ which can generate the higher, but finite order t/U expansion. 

A solitary kinetic is charge symmetric, $T_{-1,0}$, and serves as a benchmark model, along with the Hubbard and t/U models.

The general model under study is,
\begin{align}
    H = U + T_{\lambda_2,\lambda_3} + J , \label{general}
\end{align}
with $U$ and $J$ being the Hubbard and Heisenberg interactions.  Proceeding with an order parameter theory requires that $H$ be symmetrized with respect to Wick contractions. As $U$ and $J$ are familiar, consider the two-body correlated hopping in its symmetric (direct + exchange) form,
\begin{equation}
    T_2 = \frac{1}{2}\sum_{ij}t_{ij} \Big[ \rho_{ij} (\rho_i+\rho_j) - \bm{\sigma}_{ij} \cdot (\bm{\sigma}_i+\bm{\sigma}_j) \Big] \, . \label{T2}
\end{equation}
Symmetrizing three-body doublon hopping incorporates one direct, two cyclic, and three exchange diagrams,
\begin{align}
    T_3 = \frac{1}{4}\sum_{ij}t_{ij}\Big[ \rho_{ij} \big( \rho_i \rho_j +\bm{\sigma}_i \cdot \bm{\sigma}_j - |\rho_{ij}|^2 + |\bm{\sigma}_{ij}|^2 \big) \nonumber \\
    -\bm{\sigma}_{ij} \cdot (\rho_i \bm{\sigma}_j+\rho_j \bm{\sigma}_i) + i\bm{\sigma}_{ij} \cdot \bm{\sigma}_i \times \bm{\sigma}_j         \Big] \,. \label{T3}
\end{align}
The action of $T_3$ is baroque; a single Wick contraction generates an effective two-body interaction that renormalizes density interactions, spin exchange, pair hopping, spin-current interactions, correlated hopping, and a Dzyaloshinskii–Moriya interaction. Much of this action plays a formative role in the coplanar spiral state of section \ref{sec4}. First, consider the simpler collinear antiferromagnet.

Under the Hubbard-commuting constraint, $\lambda_2=-1$, and at half-filling, $\nu=1$, the general model \eqref{general} contracted in the local Néel state generates momentum independent gaps from $U$ and $J$, and quasiparticle dispersion in the comoving frame from $T,T_2$, and $T_3$,
\begin{align}
    &(t_{ij}^{\perp})^\pm = 0, \\
    &(t_{ij}^{\parallel})^+ = t_{ij}^\parallel, \\
    &(t_{ij}^{\parallel})^- = t_{ij}^\parallel (\lambda_3-1)  
\end{align}
in which notationally, the underlying band segregates into inter- and intra-sublattice elements, $\{t_{ij}\} =\{t_{ij}^\perp\} \cup \{t_{ij}^\parallel\}$. The absence of e.g. nearest-neighbor hopping $t$ from quasiparticle energies is accounted for in the cancellation of bare band structure by the density part of \eqref{T2}, and the nullification of the spin part of \eqref{T2} as $\langle\bm{\sigma}_i +\bm{\sigma}_j\rangle =0$ for nearest-neighbors $i,j$, while in the three-body kinetic \eqref{T3}, $\langle  \rho_i \rho_j + \bm{\sigma}_i\cdot\bm{\sigma}_j   \rangle=0$. The comoving dispersion is charge symmetric when $\lambda_3=0$, inheriting from the global $SU(2)$ charge symmetry of $T-T_2$, is antisymmetric when $\lambda_3=2$ (t/U model), and is asymmetric otherwise.
\begin{figure*}
    \centering
    \includegraphics[width=0.176\textwidth]{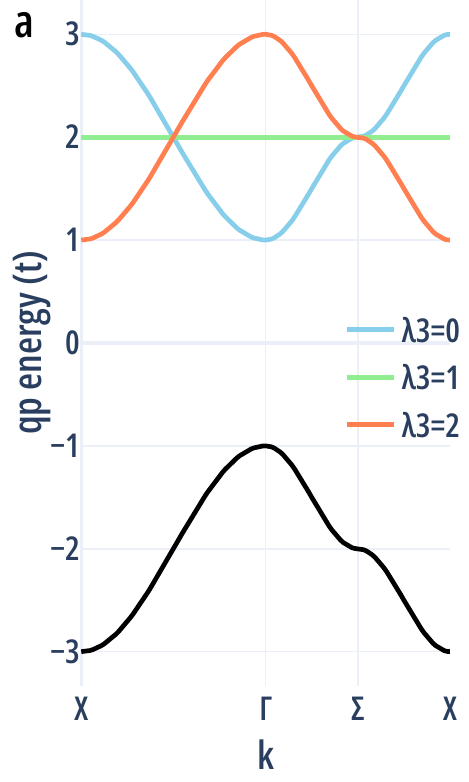}
    \includegraphics[width=0.176\textwidth]{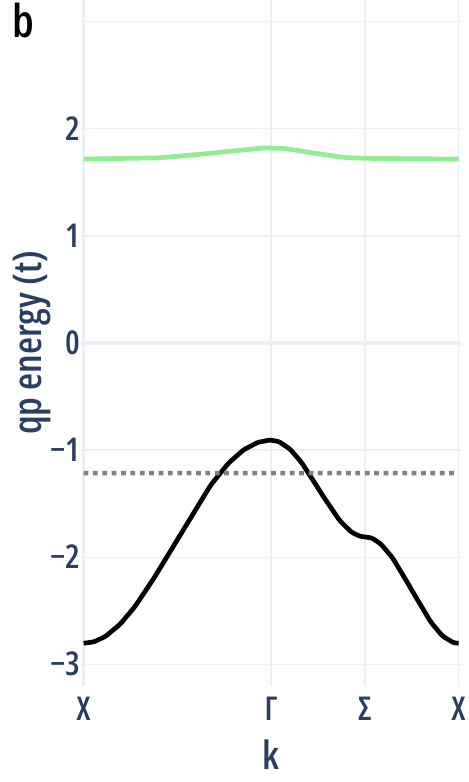}
    \includegraphics[width=0.206\textwidth]{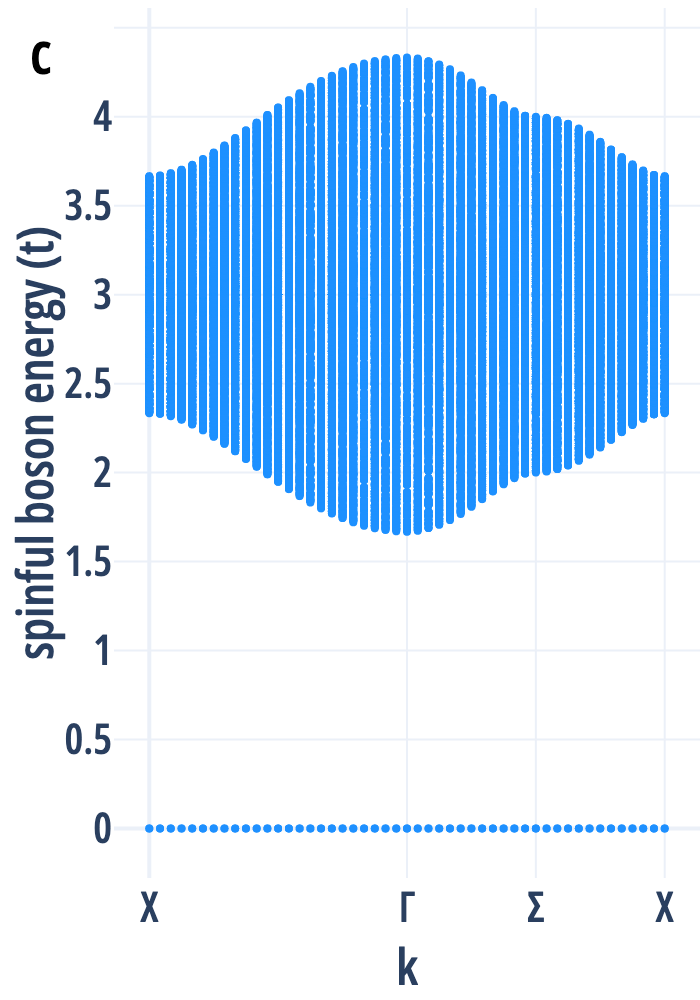}
    \includegraphics[width=0.206\textwidth]{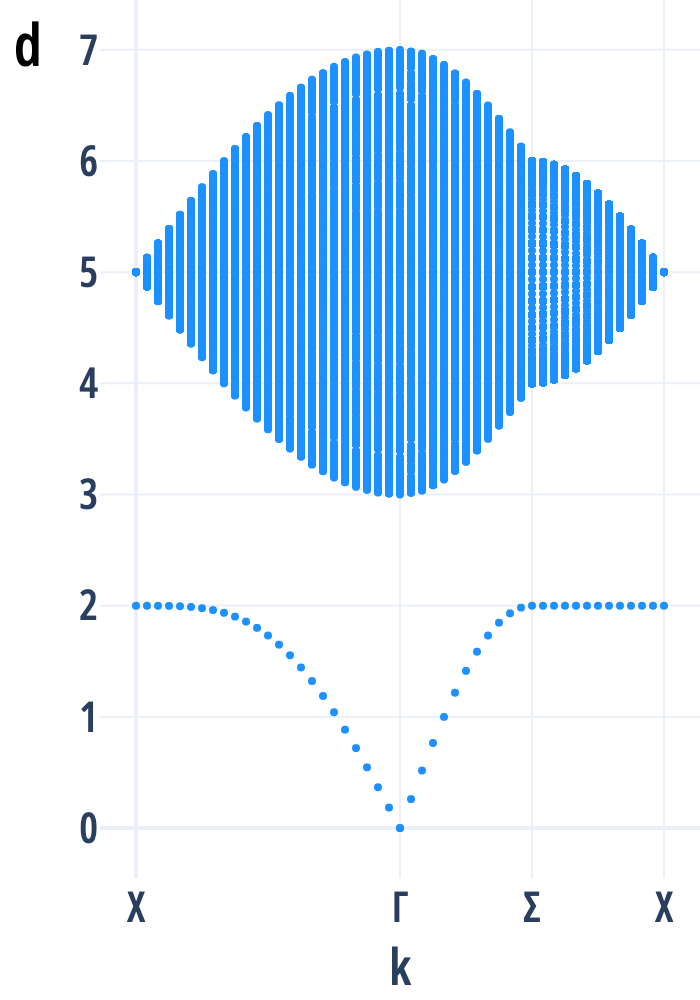}
    \includegraphics[width=0.206\textwidth]{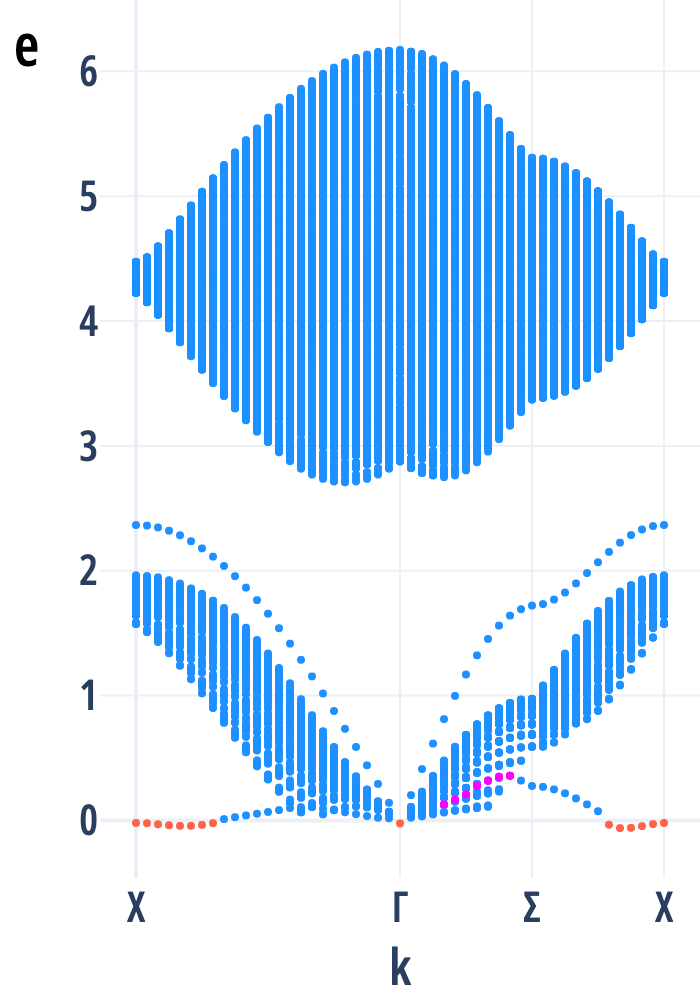}
    \caption{(\textbf{a}) Quasiparticle bands of the Néel state at $\nu=1$ for three Hubbard-commuting kinetics, $T_{-1,0}=T_h-T_d$, $T_{-1,1}=T_h$, $T_{-1,2}=T_h+T_d$, with $U=4,J=0$. (\textbf{b}) Hole doping brings dispersion to the conduction band for $\lambda_3=1$, departing from the exact doublon localization of the interacting model. (\textbf{c}) A zero-energy, flat, spin flip boson emerges from Hubbard-commuting kinetics, here $\lambda_3=2/3$ (\textbf{d}) Adding $J=1$ recovers the Holstein-Primakoff spin-wave of the Heisenberg model, here $\lambda_3=0$ (\textbf{e}) Away from half-filling, $\nu=0.93$, with $J=1/4$, the lower Hubbard band drives the primary spin-wave into the usual albeit more agressive form of spiral instability. More distinctively for tight dynamics, $\lambda_2 \sim -1$, a second spin wave is made out of the LHB, contrasting with the primary, which is almost entirely of the UHB. }
    \label{fig:2}
\end{figure*}
Néel quasiparticle energies for conduction/valence bands are $E_{\bm{k},\tau}^\pm = \pm(u/2+J) +\epsilon^\parallel_\mp(\bm{k})$, where $\epsilon^\parallel_\pm(\bm{k})$ are Fourier components of $(t_{ij}^\parallel)^\pm$, and $\tau$ refers to sublattice. Metastability of the fully polarized Néel state requires the quasiparticle bands to be fully gapped, in analogy with simple Stoner ferromagnetism.  In a sense, the mean-field theory of the classical Mott insulator manifold, the $2^M$ spin states that are degenerate when $\lambda_2=-1$ and $J=0$, is a generalization of Stoner ferromagnetism to arbitrary textures, and Mott exclusion is a generalization of Pauli exclusion. 

It is expected that all Mott insulators are unstable when the momentum-independent gap is lowered beneath quasiparticle bandwidth, and it seems generically possible that all classical states other than the pure MIs are technically unstable against the fluctuations of a fundamentally quantum dynamic \eqref{general}. However, many states, like the spiral magnet, inhomogeneous stripes, and crystals, can at least be found as converged solutions of the iterative Hartree-Fock equation. Ironically, the ferromagnet, whose essence is leveraged by Hubbard-commuting dynamics to effect Mott localization, appears to be the least stable phase; the HF algorithm applied to the $\bm{q}=0$ sector ends in a limit cycle when $u$ is less than the noninteracting bandwidth. This point is expanded upon in section \ref{sec4}. The general mean-field equations of the antiferromagnet (applicable away from $\nu=1$) can be distilled from those of the spiral in section \ref{sec4} by setting $\bm{q}=\bm{\pi}$.

The essence of localization from itinerancy is captured by a defining property of pure Mott insulators, their strict charge incompressibility, $\chi_c(\bm{q},i\omega)=0$.

\section{Emergence of the flip boson\label{sec3}}
The second, but less trivial defining quality of classical Mott insulators (for $J=0$) concerns their spin susceptibility,
\begin{equation}
    \chi_s(\bm{q},i\omega)=1/i\omega . \label{chispin}
\end{equation}
This statement is violated in the static construction, requiring time-dependent mean-field theory to elucidate. In that framework, it corresponds to the emergence of a zero energy boson for all $\bm{q}$. Physically, it reflects the energetic freedom to flip any spin, in any of the pure MIs, owing to their exact degeneracy. The satisfaction of \eqref{chispin} is a technical prerequisite of the emergence of linear spin wave theory from time-dependent Hartree-Fock when $J \ne 0$.

The central device of TDHF is the boson Hamiltonian, which counts and scatters particle-hole excitations of the HF ground state. Generally, the boson Hamiltonian has off-diagonal, Bogoliubov structure in the space of excitations and de-excitations, rendering a quantum correction to the ground state. When $\lambda_2=-1$ and $J=0$, the spin states are exact eigenstates of \eqref{general}, so there is no off-diagonal coupling. In the space of spinful ($S^z=-\hbar\tau$) excitations,
\begin{align}
&H^{(\bm{q})}_{boson}=\sum_{\bm{k}\bm{p},\tau} \big(V_{\bm{k}\bm{p}}+\delta_{\bm{k},\bm{p}}(E_{\bm{k}+\bm{q}}^+-E_{\bm{k}}^-)\big) q^\dagger_{\bm{k} \tau} q_{\bm{p} \tau}, \nonumber \\
&V_{\bm{k}\bm{p}}=\textstyle\frac{2}{M} [-u+\epsilon_\parallel(\bm{k})+\epsilon_\parallel(\bm{p}) \nonumber \\
&\;\;\;\;\;\;\;\;\;\;\;\;\;\;\;\;\;\;+(1-\lambda_3)(\epsilon_\parallel(\bm{k}+\bm{q})+\epsilon_\parallel(\bm{p}+\bm{q})) ], \nonumber \\
&q^\dagger_{\bm{k} \tau} = \Psi^\dagger_{\bm{k}+\bm{q} \tau +} \Psi_{\bm{k} \tau -} ,
\end{align}
taking note that momenta are restricted to the magnetic Brillouin zone. From the periodicity of $\epsilon_\parallel$ within the magnetic BZ, it follows that $\sum_{\bm{p}} \epsilon_\parallel(\bm{p+s})=0$, and therefore the uniform sum over $V_{\bm{kp}}$ depends only on $\bm{k}$, and $\bm{q}$, particularly $\sum_{\bm{p}}V_{\bm{kp}} = -(E_{\bm{k}+\bm{q}}^+-E_{\bm{k}}^-) $.

Therefore, a uniform superposition of itinerant particle-hole pairs, corresponding to a purely local spin flip, is a zero energy eigenstate of $H_{boson}$,
\begin{align}
    &H^{\bm{q}}_{boson}\, \Omega^\dagger_{\bm{q}\tau} \,|Neel\rangle = 0, \;\;\;\;\;\;\forall \bm{q},\tau , \nonumber\\
    &\Omega^\dagger_{\bm{q}\tau} =\sqrt{2/M} \sum_{\bm{k}} q^\dagger_{\bm{k} \tau}
\end{align}
The matrix elements appearing in the transverse susceptibility are $\langle n,\tau | \bm{S}^{+(-)}_{\bm{q}} | Neel \rangle =\delta_{n,\Omega}\delta_{\tau, B(A)}$, leading to \eqref{chispin}. The simple property of \eqref{chispin} expectedly extends to all spin states. A slightly more ornate example, the canted antiferromagnet, is presented in Appendix.

When spin-exchange is added to the mix, the effect on the static theory is to simply augment the gap, $u \to 
 u+2J_\perp(0)-2J_\parallel(0)$, where $J_{\perp,\parallel}(\bm{q})=\frac{1}{2}\sum_{\bm{\delta}}J^{\perp,\parallel}_{\delta} e^{i \bm{q}\,\bm{\delta}}$. Since the Neel state is not an eigenstate of spin-exchange, the Heisenberg term adds Bogoliubov structure to the boson Hamiltonian. The diagonal scattering term amends, $V_{\bm{kp}} \to V_{\bm{kp}} + 2J_\parallel(\bm{q}) - 2J_\parallel(\bm{k-p})$, and an off-diagonal term coupling excitations and de-excitations on opposing sublattices is generated, $4J_\perp(\bm{q})\sum_{\bm{kp},\tau} q^\dagger_{\bm{k} \tau}(-q)^\dagger_{\bm{p} -\tau} $. As before, the uniform sum over $\bm{p}$ voids $J_\parallel(\bm{k}-\bm{p})$, and since the off-diagonal term is independent of $\bm{k},\bm{p}$, i.e. is a matrix of ones, the flip boson wavefunction as a normalized vector of ones is an eigenstate with eigenvalue $1$. Consequently, as long as the flip boson is a proper eigenstate of the underlying $J=0$ boson Hamiltonian, the addition of spin-exchange $J$ leaves the dynamics closed in the space of flip excitations and de-excitations, leading to the pristine emergence of standard linear spin wave theory, with the dispersive Nambu-Goldstone mode, $E_{NG}=2((J_\perp(0)-J_\parallel(0)+J_\parallel(\bm{q}))^2 - J_\perp(\bm{q})^2)^{1/2}$. It should be noted that the flip boson itself emerges from local rather than global symmetry, generally violates the counting rules\cite{PhysRevLett.108.251602}, so cannot be regarded as an NG mode, but rather as primitive and atavistic.

 The fluctuation theory is not analytically tractable outside of the pure Mott insulator, requiring numerically exact diagonalization of the boson Hamiltonian for $\lambda_2 \ne -1$ or for $\nu \ne 1$. However, a qualitative observation can be made of infinitesimal doping. The major change is activation of the inter-sublattice part of correlated hopping $T_2$, which couples flip boson excitations to intraband (lower Hubbard band) excitations and de-excitations. Since nearest neighbor $t$ is the largest energy scale after $U$, dominating over quasiparticle bandwidth $\sim t'$, the coupling is so large that the \emph{diagonal} block of $H_{boson}$ contains negative eigenvalues, seeming to violate causal requirements of response functions. Addition of sufficiently finite $J$ seems to ameliorate this problem, leading to a strong form of the usual spiral instability.

\begin{figure*}
    \centering
    \includegraphics[scale=0.25]{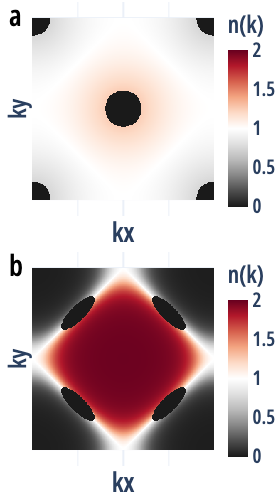}  
    \includegraphics[scale=0.37]{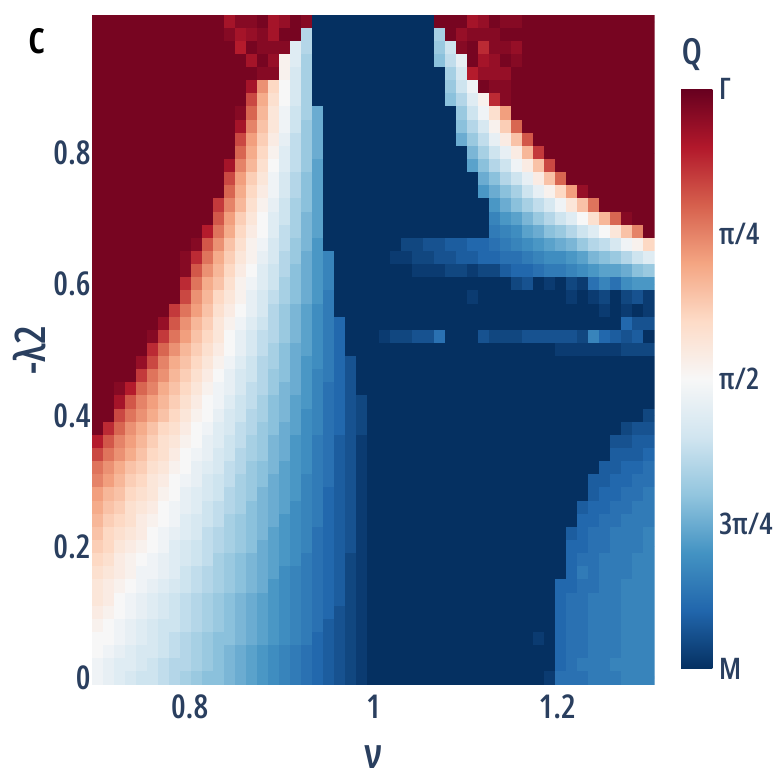}
    \includegraphics[scale=0.37]{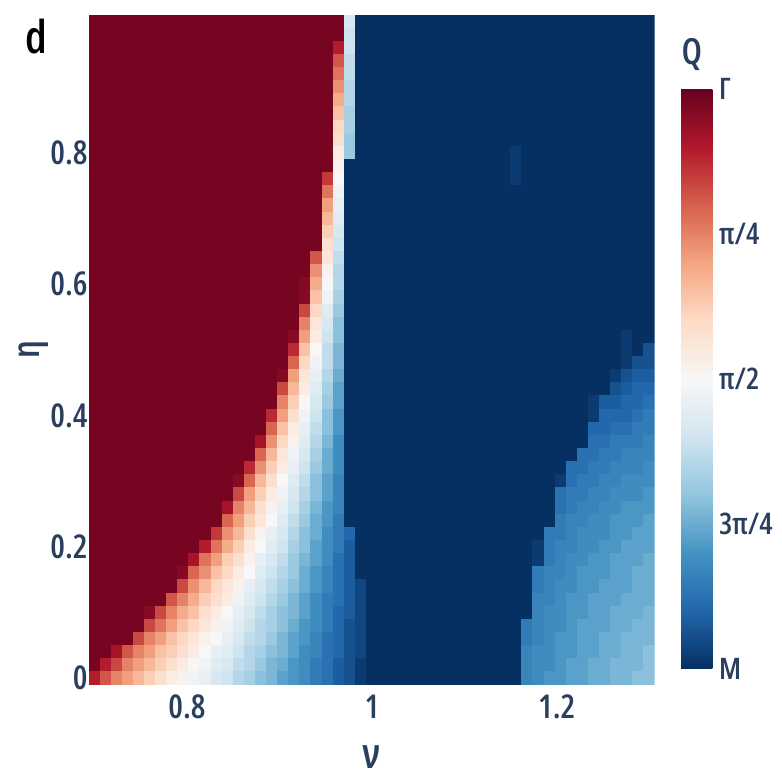}
    \includegraphics[scale=0.37]{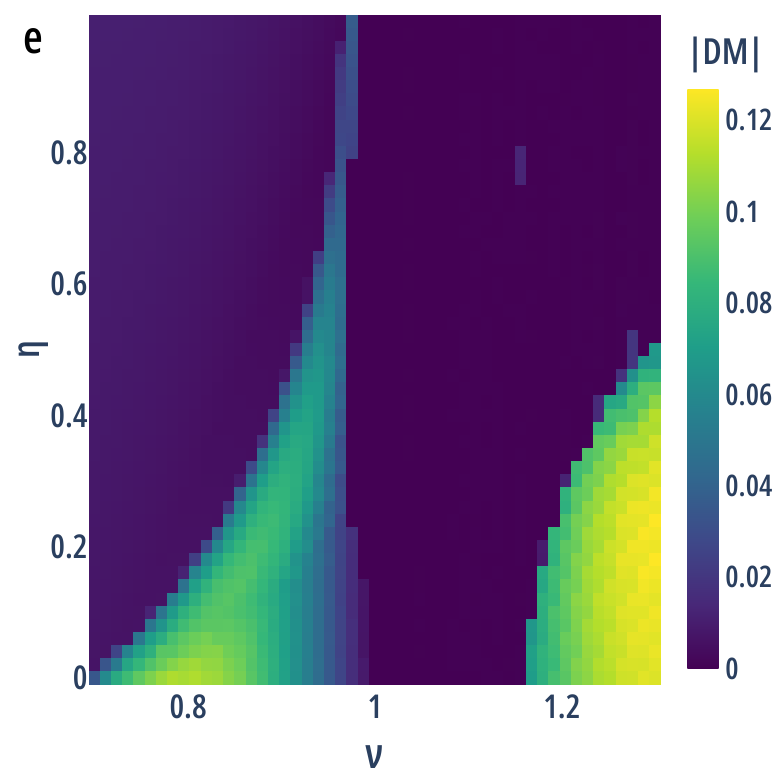}
    \caption{The transition from a gapped antiferromagnet with four nodal pockets to an AFM with two pockets $(\Gamma,M)$ is induced in the Hubbard model by increasing $U$. Beyond their role in shaping charge asymmetry of the general dynamic, $\lambda_2$ and $\eta$ function as tightness parameters, effecting the same Lifshitz transition at constant $U$. (\textbf{a}) $\eta=1, U=3, J=0$ (\textbf{b}) $\eta=0, U=3, J=0$. Bandwidths decrease as $\lambda_2 \to-1, \eta\to 1$, and become highly spin dependent, so the AFM quickly loses ground to various FM phases that are kinetically advantageous. This transition is facilitated by a piece of $T_3$ that generates ferromagnetic spin-exchange in proportion to bond kinetic energy. (\textbf{c})Optimal $\mathcal{Q}_{spiral}$ along the line from $\Gamma$ to $M$, as $\lambda_2$ moves from the HM to the symmetric model, with $U=6,J=1/3$. (\textbf{d}) Along the line of antisymmetric kinetics, from HM to t/U, $U=10,J=1/4$. (\textbf{e}) Outside of $\mathcal{Q}=0,\bm{\pi}$, inversion symmetry is absent and a spin current condenses (DM), which in turn generates a Dzyaloshinskii–Moriya interaction from $T_3$, as in the last term of \eqref{T3}}
    \label{fig:spiral}
\end{figure*}

\section{Charge asymmetry of the spin spiral\label{sec4}}

Fluctuations and instabilities of the antiferromagnet are particularly strong near Hubbard-commuting dynamics, motivating explication of spiral order instead. For simplicity, the coplanar rather than the conical spiral is considered, but the self-energy is unavoidably baroque in its inheritance from $T_3$. There are various order parameter dependent adjustments to underlying dispersion, and most notably a Dzyalshinkii-Moriya interaction is generated from the condensation of inversion-odd, time-reversal-even, spin current. Solution of the spiral from the general dynamic \eqref{general} shows that its pitch $1/\mathcal{Q}$ as a function of charge density is tuneable by the underlying band $t_{ij}$ and the overlying kinetic couplings $\lambda_2,\lambda_3$. The most novel result is the situation of a spin-delocalized kinetic ferromagnet in general proximity to the symmetric model, accessible at $\nu=1$ for small values of $U$ and at finite doping for intermediate values.

Taking note of shorthand, $\cos^{\bm{\delta}}_{\bm{a},\bm{b}}:=(\cos(\bm{a}\cdot\bm{\delta})+\cos(\bm{b}\cdot\bm{\delta}) )/2 $, and $\sin^{\bm{\delta}}_{\bm{a},\bm{b}}:=(\sin(\bm{a}\cdot\bm{\delta})+\sin(\bm{b}\cdot\bm{\delta}) )/2 $, a quasiparticle Hamiltonian drives the coplanar spiral,
\begin{align}
    H_{mf} &= \sum_{\bm{k},\alpha} \Psi^\dagger_{\bm{k}}(h^\alpha_{\bm{k}} \sigma^\alpha )  \Psi_{\bm{k}} , \;\;\;\;
    \Psi^\dagger_{\bm{k}} = (c^\dagger_{\bm{k+q}\uparrow}, c^\dagger_{\bm{k}\downarrow}) , \label{Hspiral}\\
    h^0_{\bm{k}} &= (\tilde{\epsilon}_{\bm{k}}+\tilde{\epsilon}_{\bm{k+q}})/2+(\xi_{\bm{k}}-\xi_{\bm{k+q}})/2 -\tilde{\mu} \nonumber\\
    h^3_{\bm{k}} &= (\tilde{\epsilon}_{\bm{k}}-\tilde{\epsilon}_{\bm{k+q}})/2+(\xi_{\bm{k}}+\xi_{\bm{k+q}})/2, \nonumber\\
    h^1_{\bm{k}}-i h^2_{\bm{k}} &= \Delta+\Delta_{\bm{k}} , \nonumber
\end{align}
with energies $E^{\pm}_{\bm{k}} = h^0_{\bm{k}} \pm ((h^3_{\bm{k}})^2+|\Delta+\Delta_{\bm{k}}|^2)^{1/2}$ that depend on renormalized band energies,
\begin{align}
    \tilde{\epsilon}_{\bm{k}} &= \sum_{\bm{\delta}}  \cos^{\bm{\delta}}_{\bm{k},\bm{k}} \Big(2 t_{\delta} (1+\lambda_2 \nu +\lambda_3 \nu^2/4 ) -(J_{\delta} \mathsf{b}_{\bm{\delta}}/2) \nonumber\\
    &+ 2 t_{\delta} \lambda_3 (-3\mathsf{b}_{\bm{\delta}}^2/4 + |\mathsf{m}_0|^2 \cos^{\bm{\delta}}_{\bm{q},\bm{q}} + |\mathsf{m}_{\bm{\delta}}|^2 +{\mathsf{j}^z_{\bm{\delta}}}^2/4 )\Big),\\
    \xi_{\bm{k}} &=\sum_{\bm{\delta}}  \sin^{\bm{\delta}}_{\bm{k},\bm{k}} \Big(2 t_{\delta} \lambda_3( \mathsf{j}^z_{\bm{\delta}} \mathsf{b}_{\bm{\delta}}/2 + |\mathsf{m}_{0}|^2 \sin^{\bm{\delta}}_{\bm{q},\bm{q}}  )  +J_\delta \,\mathsf{j}^z_{\bm{\delta}} /2 \Big),
\end{align}
chemical potential,
\begin{align}
    \tilde{\mu} &= \mu - u/2(\nu-1) -\sum_{\bm{\delta}}2 t_{\delta} \mathsf{b}_{\bm{\delta}}(\lambda_2+\lambda_3 \nu /2)\nonumber\\
    &+\sum_{\bm{\delta}} J_{\delta} \nu/2 -\lambda_3 \sum_{\bm{\delta}} 2 t_{\delta}(  \mathsf{m}^*_0 \mathsf{m}_{\bm{\delta}} +  \mathsf{m}_0 \mathsf{m}^*_{\bm{\delta}}),
\end{align}
and magnetization gaps,
\begin{align}
    \Delta^*&=   -u \mathsf{m}_0 + \mathsf{m}_0\sum_{\bm{\delta}} J_{\delta} \cos^{\bm{\delta}}_{\bm{q},\bm{q}} -2 \lambda_2 \sum_{\bm{\delta}} 2 t_{\delta}\mathsf{m}_{\bm{\delta}}\nonumber\\
    &+\lambda_3\sum_{\bm{\delta}} 2 t_{\delta}\mathsf{m}_0(\mathsf{b}_{\bm{\delta}}\cos^{\bm{\delta}}_{\bm{q},\bm{q}}+\mathsf{j}^z_{\bm{\delta}}\sin^{\bm{\delta}}_{\bm{q},\bm{q}}) -\lambda_3 \sum_{\bm{\delta}} 2 t_{\delta}  \nu \mathsf{m}_{\bm{\delta}}  \\
    \Delta^*_{\bm{k}} &= -2 \lambda_2 \mathsf{m}_0 \sum_{\bm{\delta}} 2 t_{\delta} \cos^{\bm{\delta}}_{\bm{k+q},\bm{k}} +\sum_{\bm{\delta}}J_{\delta} \cos^{\bm{\delta}}_{\bm{k},\bm{k}}\mathsf{m}_{\bm{\delta}} \nonumber\\
    &+\lambda_3\sum_{\bm{\delta}} 2 t_{\delta}\cos^{\bm{\delta}}_{\bm{k+q},\bm{k}}(\mathsf{m}_{\bm{\delta}}\mathsf{b}_{\bm{\delta}}  -  \nu\mathsf{m}_0 ).
\end{align}

\begin{figure}
    \centering
    \includegraphics[scale=0.41]{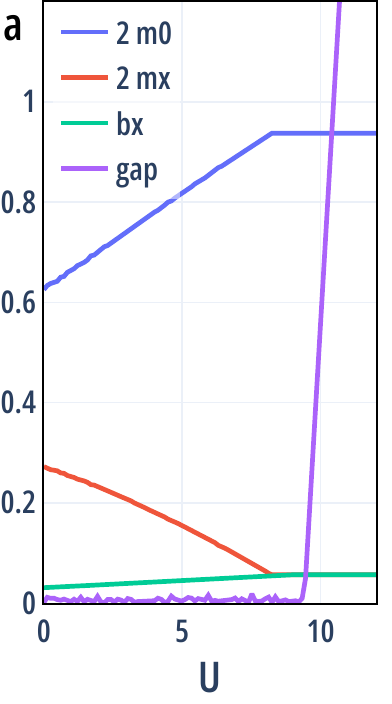}
    \includegraphics[scale=0.41]{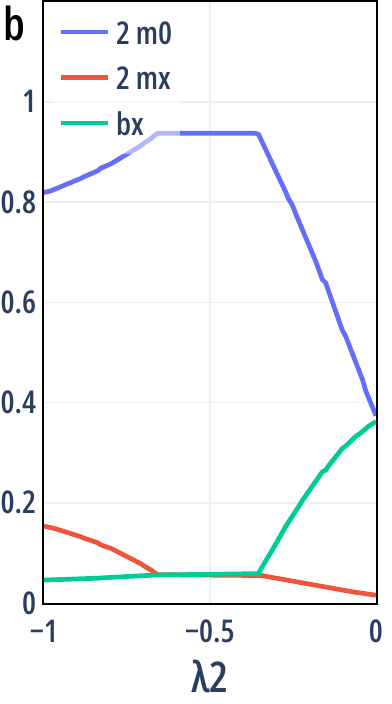}
    \includegraphics[scale=0.28]{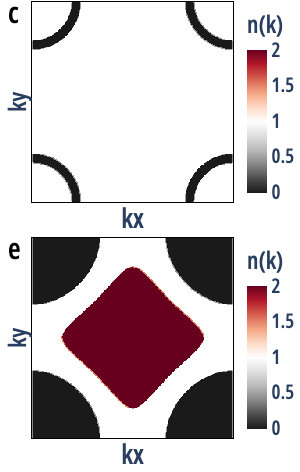}
    \caption{Lifshitz transitions of ferromagnetic phases. (\textbf{a}) Increasing $U$ alongside the symmetric kinetic $T_{-1,0}$ transitions out of a doped kinetic FM (kFM), to a gapless polarized, then to a gapped Stoner metal. (\textbf{b}) Tuning from the symmetric to Hubbard model  while holding $U=5$, from a doped kFM, to a gapless polarized, then to a gapless unpolarized FM. Notice how the essence of the kFM is domination of spin over charge delocalization, a hallmark of charge symmetry. Momentum occupancy in (\textbf{c}) the doped kFM and (\textbf{d}) unpolarized Stoner metal of the HM. Technically, the kFM is not a ferromagnet, since it requires a modicum of inversion breaking, $\bm{q}=2\bm{\pi}/L$, to stabilize.}
    \label{fig:4}
\end{figure}

These in turn depend on bond-resolved charge and magnetization densities, and on the spin current,
\begin{align}
    \mathsf{b}_{\bm{\delta}}&=\sum_{\bm{p}\sigma} \cos^{\bm{\delta}}_{\bm{p},\bm{p}} \,\langle  c^\dagger_{\bm{p}\sigma} c_{\bm{p}\sigma} \rangle /M, \\
    \nu&=\mathsf{b}_0 \\
    \mathsf{m}_{\bm{\delta}} &= \sum_{\bm{p}} \cos^{\bm{\delta}}_{\bm{p},\bm{p+q}} \,\langle  \Psi^\dagger_{\bm{p}} \sigma^{-}  \Psi_{\bm{p}} \rangle /M, \\
    \mathsf{j}^z_{\bm{\delta}} &= \sum_{\bm{p}\sigma} \sigma \sin^{\bm{\delta}}_{\bm{p},\bm{p}} \,\langle  c^\dagger_{\bm{p}\sigma} c_{\bm{p}\sigma} \rangle /M ,
\end{align}
that are measured in ground or thermal states of the quasiparticle gas driven by \eqref{Hspiral}, expressing self-consistency.  

The classical part of the energy is,
\begin{align}
    E^{(0)}/M &=u (|\mathsf{m}_0|^2 -\nu^2/4 +1/2) \nonumber \\
    +\lambda_2 &\sum_{\bm{\delta}}2 t_{\delta} \big( -\nu \mathsf{b}_{\bm{\delta}} + 2 \mathsf{m}^*_0 \mathsf{m}_{\bm{\delta}} + 2 \mathsf{m}_0 \mathsf{m}^*_{\bm{\delta}} \big) \nonumber \\
    +\lambda_3 &\sum_{\bm{\delta}}2 t_{\delta} \big( -\nu^2\mathsf{b}_{\bm{\delta}}/2 +\mathsf{b}_{\bm{\delta}}^3/2- 2\mathsf{b}_{\bm{\delta}}|\mathsf{m}_0|^2 \cos^{\bm{\delta}}_{\bm{q},\bm{q}}  \nonumber\\
    &\;\;\;\;\;\;\;\;\;\;\; -2 \mathsf{b}_{\bm{\delta}}|\mathsf{m}_{\bm{\delta}}|^2 + 2\nu \mathsf{m}^*_0 \mathsf{m}_{\bm{\delta}} + 2\nu \mathsf{m}_0 \mathsf{m}^*_{\bm{\delta}}\big) \nonumber\\
    +&\sum_{\bm{\delta}} J_{\delta} \big( \nu^2/4 + |\mathsf{b}_{\bm{\delta}}|^2/4 -|\mathsf{m}_0|^2 \cos^{\bm{\delta}}_{\bm{q},\bm{q}}-|\mathsf{m}_{\bm{\delta}}|^2\big)\nonumber\\
    +&\sum_{\bm{\delta}} \big( 2t \lambda_3 (-{\mathsf{j}^z_{\bm{\delta}}}^2\mathsf{b}_{\bm{\delta}}/2 -2\mathsf{j}^z_{\bm{\delta}} |\mathsf{m}_{0}|^2 \sin^{\bm{\delta}}_{\bm{q},\bm{q}}) - J_\delta {\mathsf{j}^z_{\bm{\delta}}}^2/4 \big)
\end{align}
while the first-order, or quasiparticle energy is,
\begin{equation}
    E^{(1)} = \sum_{\bm{k}\pm} n_f(E^{\pm}_{\bm{k}})\, E^{\pm}_{\bm{k}} .
\end{equation}
Spiral pitch $\mathcal{Q}$ in Fig \ref{fig:spiral} is determined as the best commensurate approximation to the generically incommensurate solution of $\partial_q (E^{(0)}+E^{(1)})|_{q=\mathcal{Q}}=0$, computed for a 96x96 system.

The most basic change to the constitution of the spiral from tuning kinetic interaction parameters can be seen in Fig \eqref{fig:spiral} as the dynamic tightens when $\lambda_2$
 approaches $-1$. In this Hubbard-commuting regime, the phase diagram appears as if the Hubbard interaction is very large, although in actuality it can be small, needing only to be larger than the quasiparticle bandwidth, which becomes highly sensitive to the pattern of spin order. A subtler effect is observed in the modulation of charge asymmetry, e.g. increasing $\eta$ from the HM strengthens antiferromagnetism for light hole doping, before it weakens again on approach to the t/U model. More significantly, the width of the transition from AF to FM changes quickly as $\eta$ or $\lambda_2$ depart from zero. Changing $\lambda_2$ while $\lambda_3=0$ effects a change in symmetry, and a highly asymmetric region is crossed around $\lambda_2=-0.5$.

 Qualitatively distinct phase transitions can also take hold within a definite-$\mathcal{Q}$ sector, for example between normal Stoner-type ferromagnets and a kinetic ferromagnet characterized by the delocalization of spins, depicted in Fig \ref{fig:4}. The delocalized kF also has $n_{\bm{k}}=1$ for all $\bm{k}$ at half-filling, reminiscent of Hatsugai-Kohmoto insulator. Crucially, the state is not quite a ferromagnet as it requires some amount of inversion breaking, and the mean-field theory cannot converge for $\mathcal{Q}=0$. This strongly suggests presence of a beyond classical order, like an exciton condensate of doublons and holons, or a symmetric quantum ordered metal. The interrelation of symmetry breaking and underlying entanglement is a subject of current interest in relation to models of the cuprates, e.g. in \cite{christos_model_2023}; it could be interesting to consider similar interrelations in the context of a microscopic and effective model with tunaeable charge symmetry.

\section{Discussion}
One of the major draws of cuprate phenomenology is the adjacency of exotic mystery to well understood phases. Except for the presentation of the unusual kinetic ferromagnet, the main technical results of this work are the electronic mean-field theory of pure Mott localization using kinematical interactions, and the subsequent emergence of Holstein-Primakoff theory from the time-dependent Hartree-Fock method, and are situated around the very well studied antiferromagnetic Mott insulator. These methods might find useful and contemporaneous generalization in application to Kitaev materials, where spin-dependence is already baked into the bandstructure. A natural extension in relation to the cuprates is the development of a Bogoliubov-de Gennes theory of the general dynamic \eqref{general} focused on intertwined and inhomogeneous spin, charge, and Cooper pairing order, and the driving of pair density waves. Over the past decade it was observed that pair density waves tend to exhibit local d-wave symmetry, a result seemingly inexplicable in the context of rotationally variant order. \cite{chen_identification_2022,fujita_direct_2014,choubey_atomic-scale_2020,wang_scattering_2021} This phenomenon motivates reiteration of the statement that strong selection of d-wave order follows from kinematical interactions\cite{PLAKIDA1989787}, perhaps demonstrable using the current methodology which is well suited to complex electronic order.

The allure of the exotic motivates further development of the model beyond the study of order. Measurements reveal a fast transition in the Hall resistance near the optimally doped superconductor\cite{badoux_change_2016}, inferentially related to the effective carrier density, prompting questions about Luttinger's theorem. An alternative hypothesis is that interacting kinetics lead to interacting currents, potentially effecting structural shifts in the Hall effect. An investigation along these lines would warrant careful consideration of Ward identities and full Hamiltonian effects, in analogy with treatment of superconductivity\cite{PhysRevResearch.6.013058}. 

The clearest limitation of the mean-field approximation is the absence of frequency dependence in the self-energy. A target question for higher-order perturbation theories is the extent to which the gross changes observed in quasiparticle bands persist when interactions are effectively time-dependent, and whether they take hold over broad domains of momenentum or only near Fermi or Luttinger surfaces. A potentially related set of empirical observations \cite{PhysRevLett.104.207002,PhysRevLett.105.046402} finds strong temperature and doping dependent renormalizations of fermi velocity; it would be interesting to check whether such effects emerge from more sophisticated treatments of the general dynamic \eqref{general}.

Ultimately, the goal in constructing an effective mean-field theory of Hubbard-like models aimed at the cuprates is not to elicit all behavior from a basic method, but to search for the emergence of familiar and exotic phenomena, together from a unified framework.

\begin{acknowledgments}
I wish to thank Allan MacDonald and his research group for their hospitality while this work was finished, and thank Erez Berg for a helpful discussion.
\end{acknowledgments}
\nocite{*}
\bibliography{Heisenberg}

\end{document}